\documentclass[twocolumn,twoside,slac_two]{revtex4}
\usepackage{graphicx}
\usepackage{fancyhdr}
\def\fermi{{\it Fermi \/}}

\begin{document}

\title{A search for possible dark matter subhalos as IACT targets in
  the First \emph{Fermi}-LAT Source Catalog}

\author{D. Nieto$^{\dagger}$, V. Mart\'inez, N. Mirabal, J. A. Barrio,
  K. Satalecka, S. Pardo, \& I. Lozano} \affiliation{Dpto. F\'isica At\'omica,
  Facultad CC. F\'isicas, Universidad Complutense de Madrid, E-28040,
  Spain
  ($^{\dagger}$E-mail:~\href{mailto:nieto@gae.ucm.es}{nieto@gae.ucm.es})}

\begin{abstract}
We present a systematic search for potential dark matter subhalos in our Galaxy among the 
630 unassociated sources included in the First \emph{Fermi}-LAT Source Catalog. Assuming a 
hypothetical dark matter particle that could generate observable gamma-ray photons 
beyond the \fermi energy range through self-annihilation, we look for reasonable targets 
for ground-based Imaging Atmospheric Cherenkov Telescopes at energies $E > 100$ GeV. In 
order to narrow the origin of these enigmatic sources, we look for their possible 
counterparts in other wavelengths including X-ray, radio, and optical spectroscopy. We 
find that the synergy between \fermi and Cherenkov telescopes, along with multiwavelength 
observations, could play a key role in indirect searches for dark matter. 
\end{abstract}

\maketitle

\thispagestyle{fancy}

\section{Introduction}
A gamma-ray signal in the very high energy (VHE) regime from dark
matter (DM) particle annihilation would be characterized by a very
distinctive spectral shape due to features such as lines
\citep{bertone}, and internal bremsstrahlung \citep{bring}, as well as
a characteristic cut-off at the DM particle mass. The DM
spectrum must be universal; hence a possible smoking-gun for DM would
be the detection of several gamma-ray sources, all of them sharing
identical spectra \citep{lee}.  No DM signal has been detected so far
in any of the most promising DM targets, including dSph galaxies
\citep{aleksic1}, the Galactic Center \citep{abram}, and clusters of
galaxies \citep{aleksic2}.  Yet, there might be additional regions
with high DM density.

\begin{figure}[t]
\centering
\includegraphics[width=65mm]{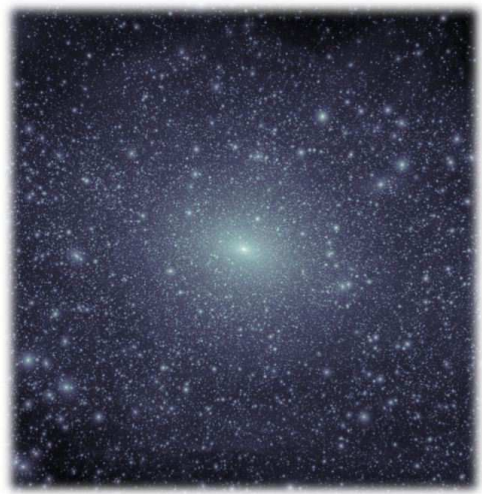}
\caption{Via Lactea II simulation of a Milky Way-sized DM halo where
  rich DM substructure emerges.  DM subhalos could
  be close enough to be detectable in the gamma-ray range. Extracted
  from~\citet{diemand}}\label{f1}
\end{figure}

High resolution simulations indicate that DM halos must exhibit a
wealth of substructure on all resolved mass scales
\citep{diemand,stadel} (see Figure~\ref{f1}). These subhalos could be
too small to have attracted enough baryonic matter to start
star-formation and would therefore be invisible to past astronomical
observations \citep{pieri} but most probably visible at HE and VHE via
annihilation of weakly interacting massive particles (WIMP). Since DM
emission is expected to be non variable in time, such hypothetical
sources would appear in the all-sky monitoring programs
\citep{kamionkowski}, and thus could be detected by the \fermi
satellite telescope as unassociated \fermi objects (UFOs) not
detected at any other wavelengths.  As mentioned above, a potential
indicator of DM detection could be a distinct cut-off close to the DM
particle mass. In the neutralino framework \citep{jung}, such a
cut-off would be likely located at energies where \fermi is not
sensitive enough \citep{amsler}. Therefore, the synergy between \fermi
and imaging atmospheric Cherenkov telescopes (IACTs) appears as a
natural way to attack this problem, since IACTs are more sensitive at
VHE.

\begin{figure*}[t]
\centering
\includegraphics[angle=-90,width=\textwidth]{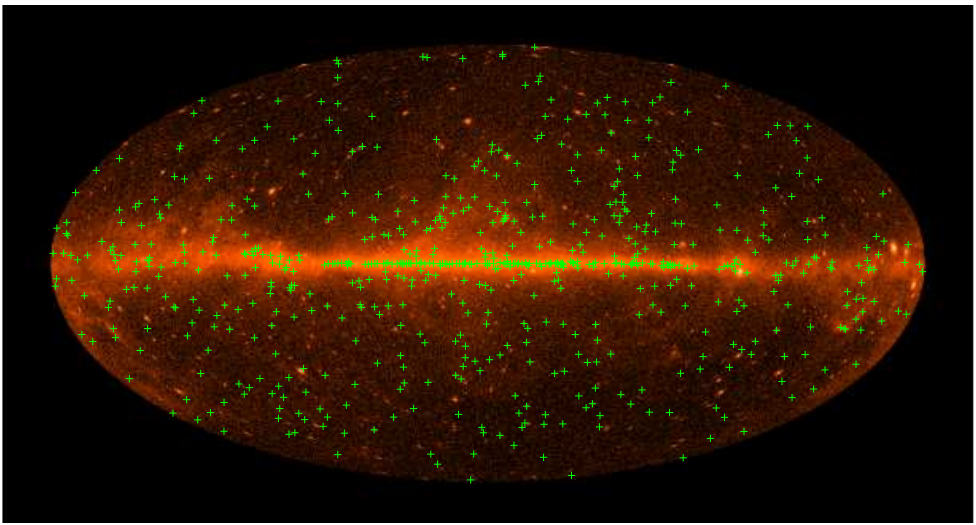}
\caption{All-sky \fermi Aitoff projection of photons above 10 GeV
  (Galactic coordinates). Green crosses indicate the nominal position
  of the 630 UFOs. As clearly seen, the bulk of
  UFOs is located in the Galactic plane.}\label{f2}
\end{figure*}

\section{Selection of Dark Matter Subhalo Candidates}
The First \emph{Fermi}-LAT catalog~\citep[1FGL][]{abdo} consists of
1451 sources with 630 unassociated with any known type of feasible
gamma-ray emitter (see Figure~\ref{f2}). The collection of possible
DM subhalo candidates out of the 1FGL starts with a selection
of UFOs based on spectral characteristics, time variability, possible
associations, and location in the sky. To qualify as a candidate, the
sources are required to meet the following criteria:

\begin{itemize}
\item A location outside the Galactic Plane.\\ 
The majority of the UFOs are located in the Galactic plane, where an
overwhelming fraction of conventional galactic objects are found
(pulsars, pulsar wind nebulae, supernova remnants, binary systems,
etc). On the other hand, the galactic DM substructures
present an homogeneous distribution in galactic
latitude~\citep{diemand,springel}. The association
algorithms are not very efficient in very crowded environments and
unassociations due to an excess of candidates are likely. On top of
that, the galactic diffuse gamma-ray background is stronger close to
the Galactic plane, making \fermi data analysis much difficult up to
the extent that the statement of detection of some faint UFOs, nearby
or within the Galactic Plane, depends on the assumed galactic
gamma-ray background model. Consequently, in case some UFOs are actual
DM clumps, the chances of ordinary object contamination in
the final selection would be much higher if low galactic coordinate
objects are considered. As a conclusion, UFOs with galactic latitudes
$|b|<10\deg$ were rejected.
\item Hardness.\\ 
The expected spectral shape from WIMP annihilation, which essentially
follow the shape of the annihilation photon yield, is hard until the
WIMP mass cut-off~\citep{cembranos}. Moreover, 1FGL sources presenting
hard spectra are more likely to be detected by IACTs.  Therefore, only
hard sources were selected, meaning that 1FGL sources with spectral
fitting power law indices $\Gamma < 2$ were considered.
\item Non variability.\\ 
The photon flux from DM annihilation must be constant over
time, thus variable sources must be rejected.  The 1FGL provides a
\emph{variability index} for each source. The corresponding light
curve is significantly different from a flat one if that index is
greater than 23.21. Therefore, sources whose \emph{variability index}
surpasses that limit were discarded.
\item Spectral behavior.\\
In the SUSY DM framework, the neutralino has a mass lower limit of
$\sim 50$ GeV \citep{jung}. Thus, the energy cut-off of its
annihilation spectrum must lay above that energy. As such, the
spectrum within the \fermi energy range must be well described by a
single power law~\citep{bertone2}.  In order to quantify departures
from a power law spectra, the 1FGL includes the so called
\emph{curvature index}. When the value of that index is greater than
11 it means that the spectrum of the source deviates from a power
law. Consequently, sources with a \emph{curvature index} surpassing
that limit were discarded.
\end{itemize}

Out of the total 630 UFOs only 93 of them fulfilled the before
mentioned criteria.

\begin{figure*}[t]
\centering
\includegraphics[width=45mm]{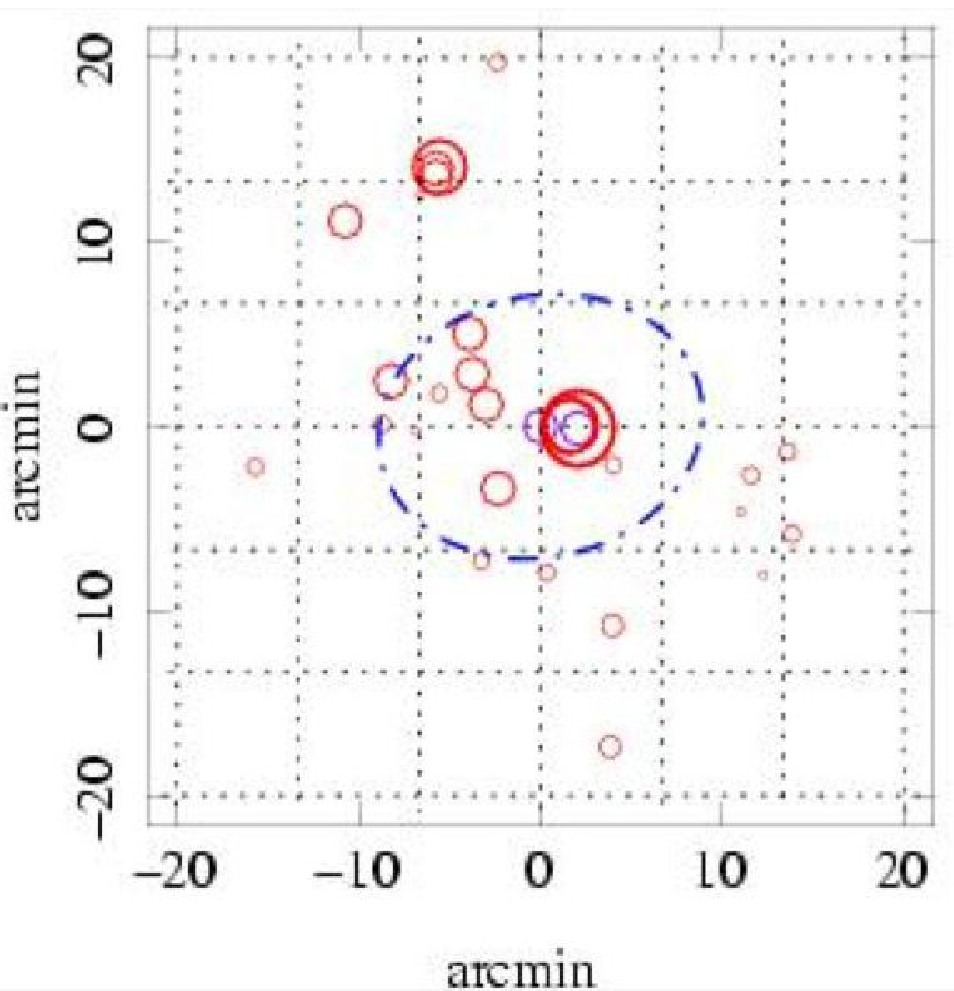}
\includegraphics[width=45mm]{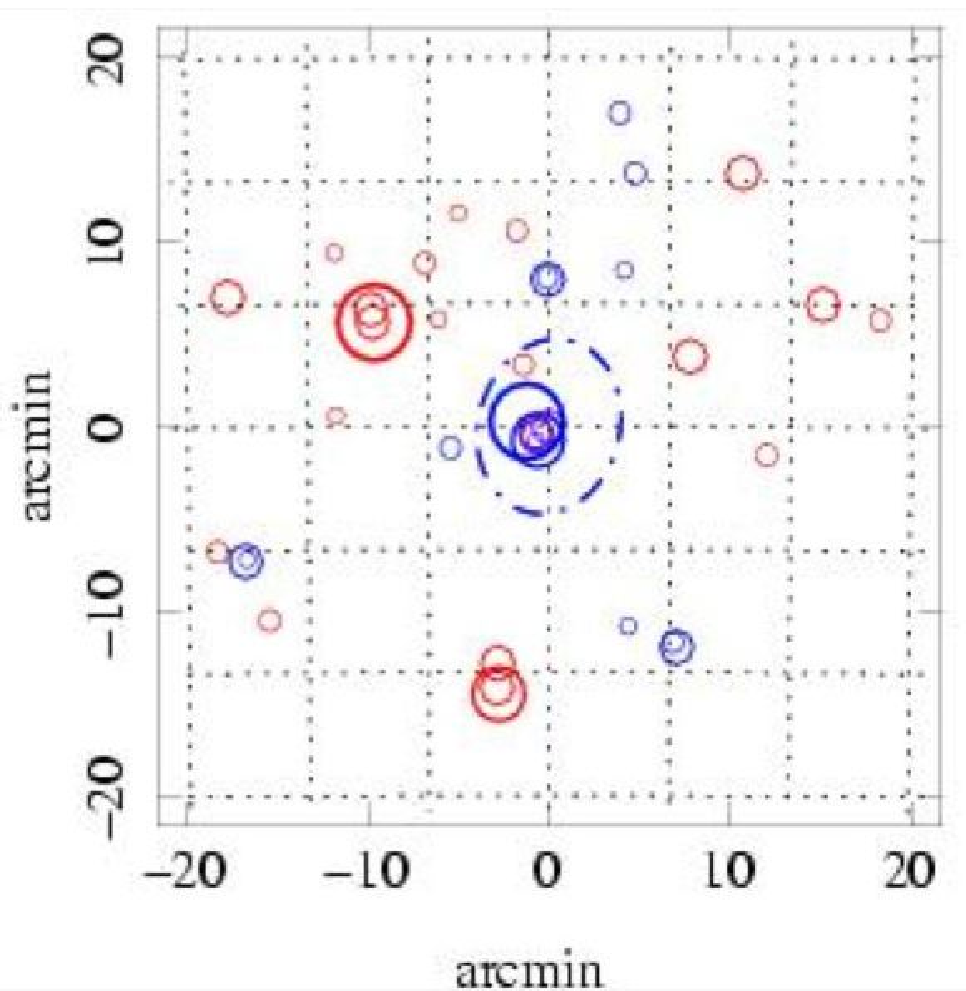}
\includegraphics[width=45mm]{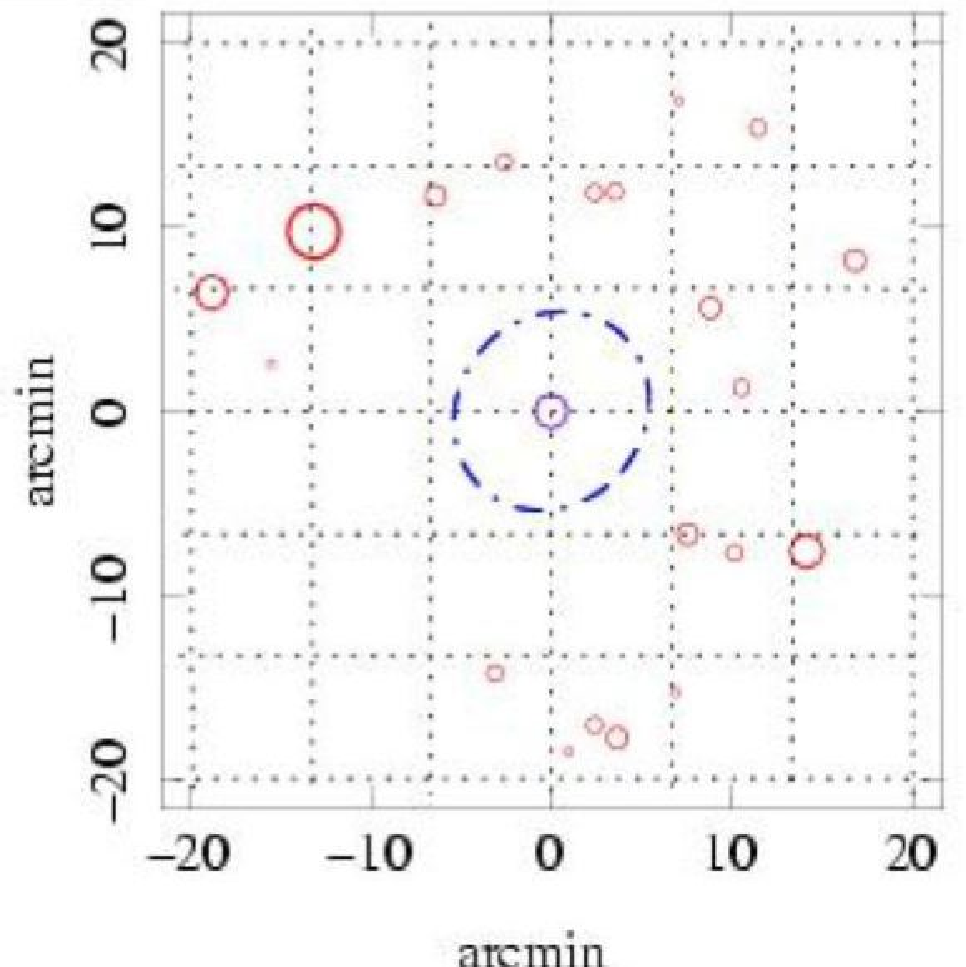}
\caption{$40^{\prime}\times40^{\prime}$ regions centered in some
  representative sample of unassociated \fermi objects out of the
  catalog selection. The purple circles are centered in \fermi nominal
  position, surrounded by a dot-dashed line representing its 95\%
  error region. Left-most and center sources were discarded since
  bright radio or X-ray sources were present within their error
  regions. The right-most field qualifies it as a candidate.  Red,
  blue, and purple circles depict radio, X-ray and gamma-ray sources
  respectively.  The maps, in equatorial coordinates, were generated
  with the ASDC Data
  Explorer~\citep{ASDCweb}.}\label{f3}
\end{figure*}

\section{Possible counterpart search}
For each candidate from the above mentioned subset of sources, we
conducted an extensive counterpart search for possible
associations. The main astronomical catalogs and mission archives were
explored around the 1FGL nominal positions with a conservative 20’
search radius, corresponding to twice the \fermi PSF at 10 GeV
\citep{burnett}.  The search, performed with the help of the NASA's
High Energy Astrophysical Archive \citep{HEASARCHweb}, scrutinized the
archives from current and past gamma-ray missions like AGILE,
INTEGRAL, CGRO, HETE-2, COS-B; X-ray missions like ROSAT,
\emph{Chandra}, XMM-\emph{Newton}, \emph{Swift}, \emph{Suzaku}, RXTE;
and radio catalogs including the NRAO VLA Sky Survey, \emph{Green
  Bank} Survey, FIRST Survey. Infrared and ultraviolet missions
archives like \emph{Spitzer}, IRAS, FUSE, and GALEX were also
considered. The purpose of this search is to discard sources whose
\fermi gamma-ray flux could be eventually attributed to an already
detected conventional source. In this way a set of unassociated
sources, i.e. sources with no potential counterparts in their \fermi
error region, was obtained. In order to illustrate the results of the
search some examples of $40^{\prime}\times40^{\prime}$ regions
centered in different UFOs are shown in Figure~\ref{f3} for both
selected and discarded sources.

After the dedicated search only 23 out of the previous 93 UFOs
surviving the catalog selection were left. \emph{Swift}-XRT data
\citep{donato} were publicly available for all these 23 sources and
were analyzed. UFOs containing X-ray sources within \fermi error
contour in \emph{Swift}-XRT data were consequently discarded. Finally,
only 10 UFOs out of the previous 23 sources, qualified as candidates.

\section{Final List}
As a final step we rank the 10 sources based on the number of
high-energy \fermi photons (E$_{\gamma}~>~$10 GeV). This number is a
crucial quantity that provides evidences for a possible extrapolation
of \fermi fluxes beyond the IACTs energy thresholds. 

\fermi data for all these 10 sources were analyzed using the latest
version of \emph{Fermi} \texttt{ScienceTools} \citep{FERMItools}. The
best suited event selection quality cuts for off-plane point source
analysis were applied by means of the \texttt{gtselect} tool, namely,
event class 3 and 4 were considered for photons below 20 GeV and class
4 beyond that energy, a maximum zenith angle cut of 105$^{\circ}$ was
applied and the latest Instrument Response Functions
(\emph{Pass6\_v3}) were considered. Regarding the time selection,
performed with the \texttt{gtmktime}, only good time intervals were
considered. On top of that, photons arriving when the satellite was
crossing the South Atlantic Anomaly were discarded as well as those
recorded at a rocking angle greater than 45$^{\circ}$. Regions of
interest-based zenith angle cuts were also applied. A circular region
corresponding to 1.5 times the \fermi PSF radius at 10 GeV
($0.15^{\circ}$) centered on the source nominal position was examined
in order to get the high-energy photons likely to have been emitted by
the source. The diffuse high-energy gamma-ray background at high
galactic latitudes is expected to be almost negligible. The background
contribution to the total number of photons extracted from the
$0.15^{\circ}$ radius region was estimated to span from $\sim0.2$ to
$\sim0.8$ photons, depending on the source.  Attending to the
estimated number of background photons in the extraction region, it is
clear that, for most of the selected UFOs, the majority of extracted
high-energy photons are unlikely to be background photons. The list of
photons per source is found in Table~\ref{table1}.

We posit that this list can serve as a source pool for follow-up
observations with IACTs. In particular, experiments such as
MAGIC~\citep{MAGICweb} hold a clear advantage based on outstanding
response at low energies (E$~<~$150 GeV) that best overlaps with the
\fermi energy range. This list could also serve to consider the
prospects with future IACT experiments.

\begin{table}[t]
\begin{center}
\caption{\emph{Fermi}-LAT photons as of February 2011. Reprocessed data (P6\_V3\_DIFFUSE)
were considered.}
\begin{tabular}{|l|c|}
\hline \textbf{Candidate} & \textbf{\emph{Fermi}-LAT photons over 10 GeV}
\\
\hline UFO I & 12.7, 14.0, 14.2, 18.2, 22.3, 23.7, 29.1, 133.5  \\
\hline UFO II & 15.6, 45.7, 20.4, 29.2, 86.8, 101.1\\
\hline UFO III & 14.5, 14.6, 22.4, 35.4, 42.5, 58.3\\
\hline UFO IV & 10.6, 24.4, 25.5, 49.2 \\
\hline UFO V & 10.0, 10.6, 12.7, 27.0\\
\hline UFO VI & 43.7, 45.4, 171.5 \\
\hline UFO VII & 15.4, 18.0, 43.1 \\
\hline UFO VIII & 18.6, 71.8\\
\hline UFO IX & 19.0, 25.0\\
\hline UFO X & 13.8, 17.0\\
\hline
\end{tabular}
\label{table1}
\end{center}
\end{table}

\section{Summary \& Outlook}
We have presented a method to select possible dark matter subhalos candidates
among unassociated \fermi objects. While there is no guarantee that the selected
candidates are \emph{bona fide} dark matter subhalos \citep{sou}, the method
presented here will help to highlight peculiar objects in the sky.

With the recently released results by \fermi, the next natural
step will be the application of the method to the 2FGL. We should then
consider the detection prospects of these sources with current IACTs
such as MAGIC and H.E.S.S.~\citep{HESSweb}, as well as with the next
generation of IACTs, namely the Cherenkov Telescope
Array~\citep{cta}.

\bigskip
\begin{acknowledgments}
This research has made use of data and/or software provided by the
High Energy Astrophysics Science Archive Research Center (HEASARC) and
the ASI Science Data Center (ASDC).  We acknowledge the Via Lactea II
project and its authors for Figure~\ref{f1}.  The authors thank the
UCM-GAE members for fruitful discussions and comments. The authors
acknowledge the support of the Spanish MICINN Consolider-Ingenio
MULTIDARK CSD2009-00064.  N.M. gratefully acknowledges support from
the Spanish MICINN through a Ram\'on y Cajal fellowship.
\end{acknowledgments}

\bigskip

\bibliography{references}

\end{document}